\input{aipcheck}

\documentclass[
    ,final            
  ]
  {aipproc}

\layoutstyle{8x11single}

 \pdfoutput=1

\usepackage{comment}

\newcommand{\enrgecoax} {{$^{\rm{enr}}$Ge-coax}}
\newcommand{\natgecoax} {{$^{\rm{nat}}$Ge-coax}}

\newcommand{\qbb}{{Q$_{\beta\beta}$}}
\newcommand{\gerda}     {{\sc Gerda}}

\newcommand{\mage}      {{\sc MaGe}}

\newcommand{\igex}     {{\sc Igex}}
\newcommand{\hdm}     {{\sc HdM}}

\def\enrgecoax	{$^{enr}$Ge-coax}
\def\qbb	 {Q$_{\beta\beta}$}
\def\ctsper      {cts/(keV$\cdot$kg$\cdot$yr)}

\def\twonu {$2\nu\beta\beta$}
\def\nonu  {$0\nu\beta\beta$}

\def\gess      {{$^{76}$Ge}}

\begin{document}

\title{Background modeling for the \gerda\ experiment}


\author{N. Becerici-Schmidt on behalf of the \gerda\ collaboration}{
  address={Max-Planck-Institut f\"ur Physik, M\"unchen, Germany }
}

\begin{abstract}
The neutrinoless double beta (\nonu) decay experiment \gerda\ at the LNGS of INFN 
has started physics data taking in November 2011. 
This paper presents an analysis aimed at understanding and modeling the observed background 
energy spectrum, which plays an essential role in searches for a rare signal like \nonu\ decay.
A very promising preliminary model has been obtained, with the systematic uncertainties 
still under study. Important information can be deduced from the model such as
the expected background and its decomposition in the signal region. 
According to the model the main background 
contributions around \qbb\ come from $^{214}$Bi, $^{228}$Th, $^{42}$K, $^{60}$Co and $\alpha$ 
emitting isotopes in the $^{226}$Ra decay chain, with a fraction depending on the assumed source positions. 
\end{abstract}

\maketitle

\section{Introduction}
Neutrinoless double beta (\nonu) decay, ($A$, $Z$) $\rightarrow$ ($A$, $Z + 2$) $+\,2e^{-}$, 
is a hypothetical process with an extremely low expected rate.  
For some even-even nuclei $\beta$ decay is energetically forbidden. They can however 
simultaneously emit two electrons and two antineutrinos via neutrino accompanied double beta 
(\twonu) decay. These nuclei can make a \nonu\ transition if lepton number 
is violated and if neutrino has a Majorana component, thus leading to physics beyond
the standard model of particle physics~\cite{0nubbReviews}. 
The expected signal of \nonu\ decay is a peak at the Q$_{\beta\beta}$ value of the decay.
The lower limits with 90\% C.L. on the \nonu\ half life 
of \gess\ are given by \hdm\ and \igex\ 
experiments as $1.9\cdot10^{25}$\,yr~\cite{klap01} and $1.6\cdot10^{25}$\,yr~\cite{igex01}, respectively.  
There is also a controversial claim of observation with a half life of 
$1.19\cdot10^{25}$\,yr~\cite{claim} from a subgroup of the \hdm\ experiment.

The GERmanium Detector Array (\gerda) experiment at the National Gran Sasso Laboratory 
(LNGS) of INFN is searching for \nonu\ decay of the \gess\ isotope~\cite{gerda}. 
The physics data taking for Phase I has started in November 2011, with the goal of testing the claim. 
The achieved background index (BI) around \qbb\ is an order of magnitude lower 
than the one of the precursor experiment \hdm. The first physics result of Phase I 
is a measurement of the half life of \twonu\ decay as $1.84^{+0.14}_{-0.10}\cdot 10^{21}$\,yr~\cite{gerda2nu}. 
Due to the superior signal-to-background ratio, a precision comparable to latest results which were obtained with a much more exposure has been achieved.  

\gerda\ follows a blind analysis strategy in Phase I; events in a 40\,keV window around \qbb\ are not 
available for analysis. The unblinding is planned for Summer 2013, when a sufficient exposure is acquired 
and the selection cuts are finalized. 
In this paper, an analysis for modeling of the observed background spectrum in Phase I 
is described and preliminary results are shown. 

\section{Experimental setup and data taking}
The \gerda\ experiment implements a novel technique by operating an array of high-purity 
germanium (HPGe) detectors directly submerged in liquid argon (LAr). 
The Phase I physics data taking has started in November 2011 with semi-coaxial p-type HPGe detectors,
eight of them enriched in $^{76}$Ge and three of them with natural abundance. 
BEGe type detectors produced for Phase II are also being tested in Phase I setup. The details of 
the \gerda\ experiment and Phase I data taking has been presented in~\cite{gerda}.

\begin{figure}[hbtp]
\includegraphics[width=0.248\columnwidth]{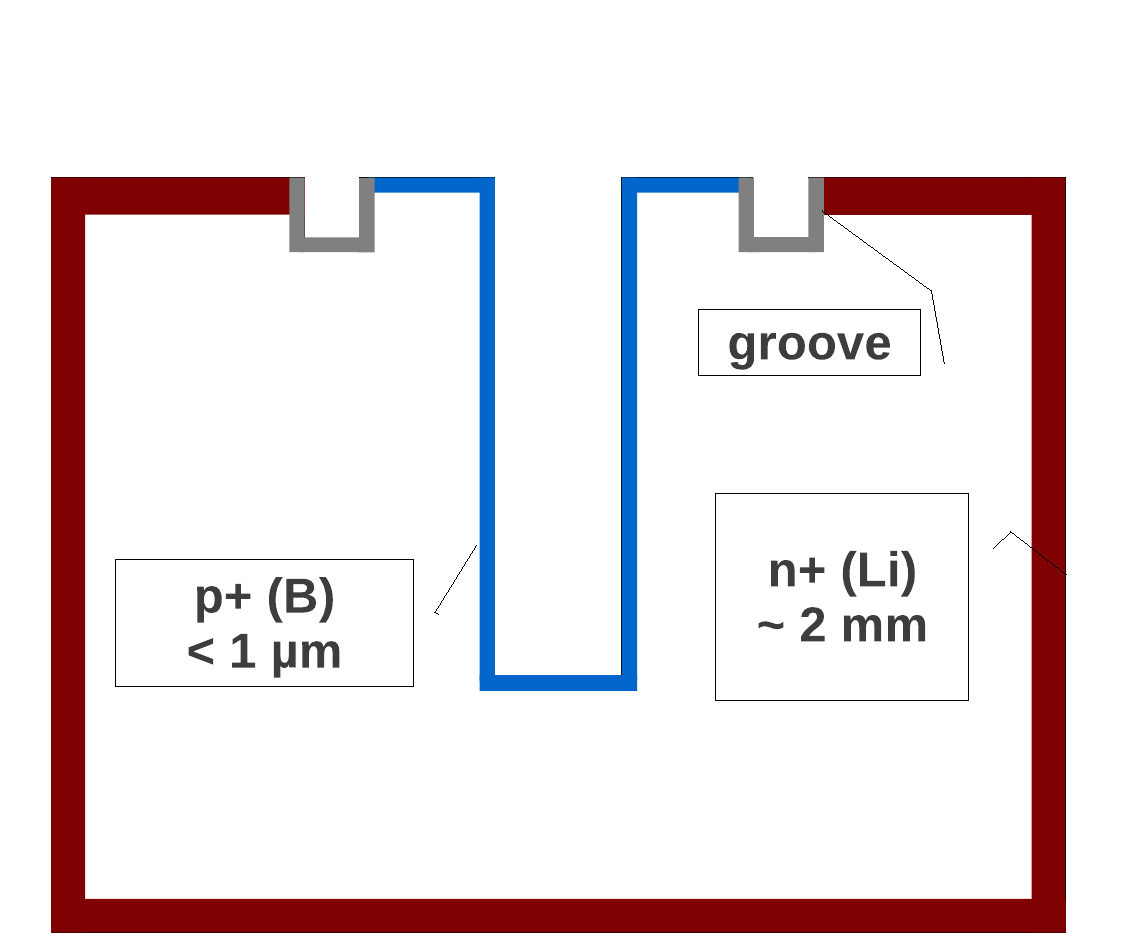}
\includegraphics[width=0.345\columnwidth]{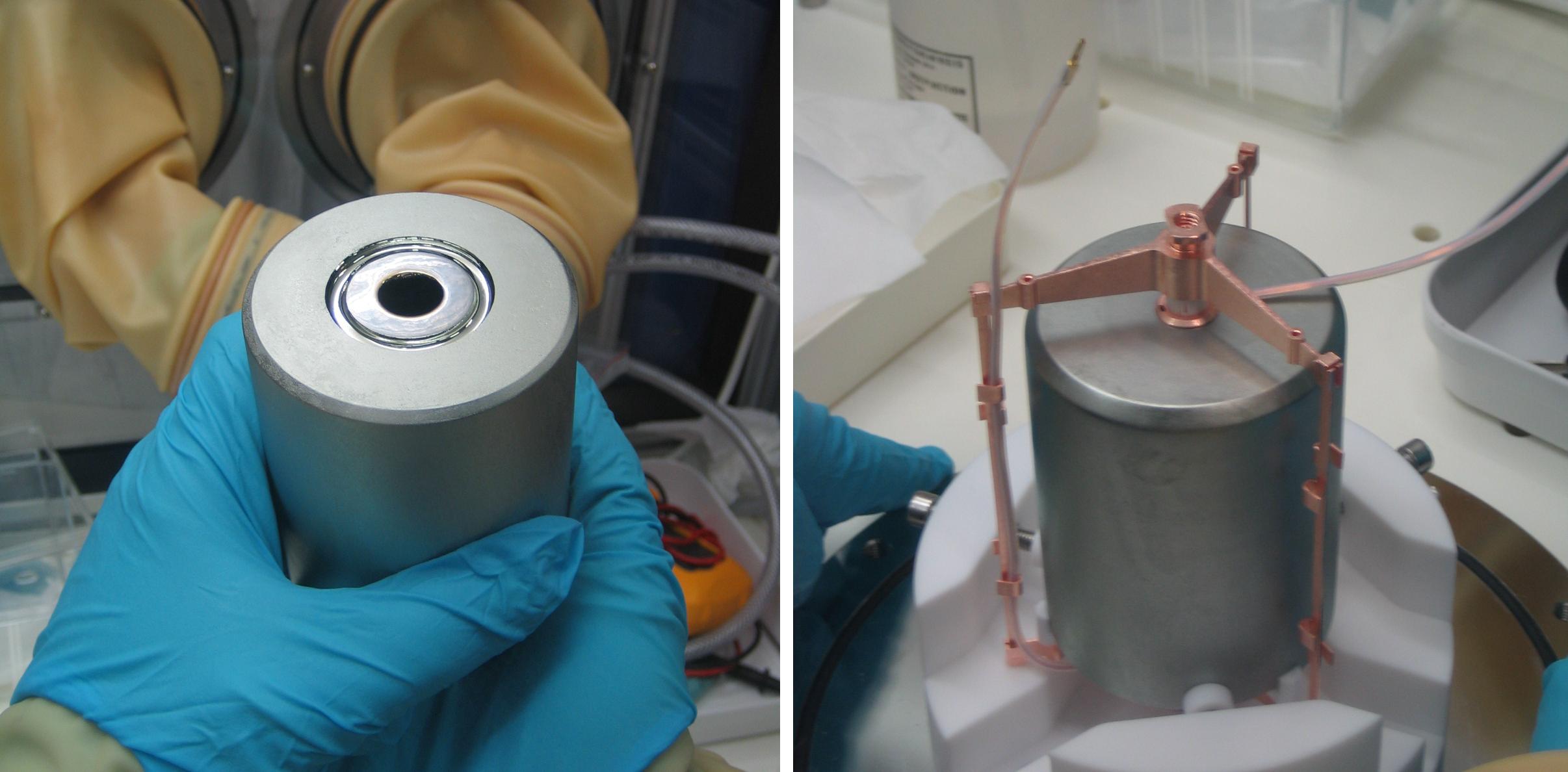}
\caption{ 
Left: Schematic drawing of a coaxial type HPGe detector.
Middle: A Phase I detector after reprocessing. Right: The detector is 
mounted upside down in its holder.
}\label{fig:coaxial}
\end{figure}

The analysis is performed on the data from coaxial detectors acquired until March 2013, 
with a total exposure of 13.65 kg$\cdot$yr for the sum enriched coaxial detectors (\enrgecoax) 
and 2.77 kg$\cdot$yr for one natural coaxial detector (\natgecoax) considered here. 
The surface of the detectors have a conductive lithium layer (n$^{+}$ contact) and a boron implanted layer (p$^{+}$ contact) 
which are separated by a groove. They form dead layers ($dl$) on the surface measured to be nearly 2\,mm for the n$^{+}$ 
and expected to be less than a $\mu m$ for the p$^{+}$ surface (see Figure~\ref{fig:coaxial}.)

\section{Analysis of the background spectrum} 
The main background sources in \gerda\ Phase I, identified by their characteristic 
$\gamma$ lines or by other features in the observed energy spectrum, are $^{60}$Co, $^{40}$K, $^{39}$Ar and $^{42}$K ($^{42}$Ar) due to LAr, 2$\nu\beta\beta$ decay of $^{76}$Ge, $^{226}$Ra ($^{232}$U-series), $^{228}$Ac and $^{228}$Th ($^{232}$Th-series). At any rate, presence of these sources in the setup was known mainly due to the screening of materials for radio purity tests or due to the LAr surrounding the bare detectors.

This section describes the analysis for modeling the observed background energy spectrum.
Firstly, the energy region above 3.5\,MeV (Q$_\beta$ of $^{42}$K) is analyzed. 
Practically no significant contribution from sources other than $\alpha$ 
decays is expected in this region. 
After obtaining an $\alpha$ model that describes the spectrum at high energies, 
a larger energy window is analyzed that includes the \qbb\ region.

\subsection{Analysis of the $\alpha$-induced background}
\begin{figure}[htpb]
\includegraphics[width=0.56\columnwidth]{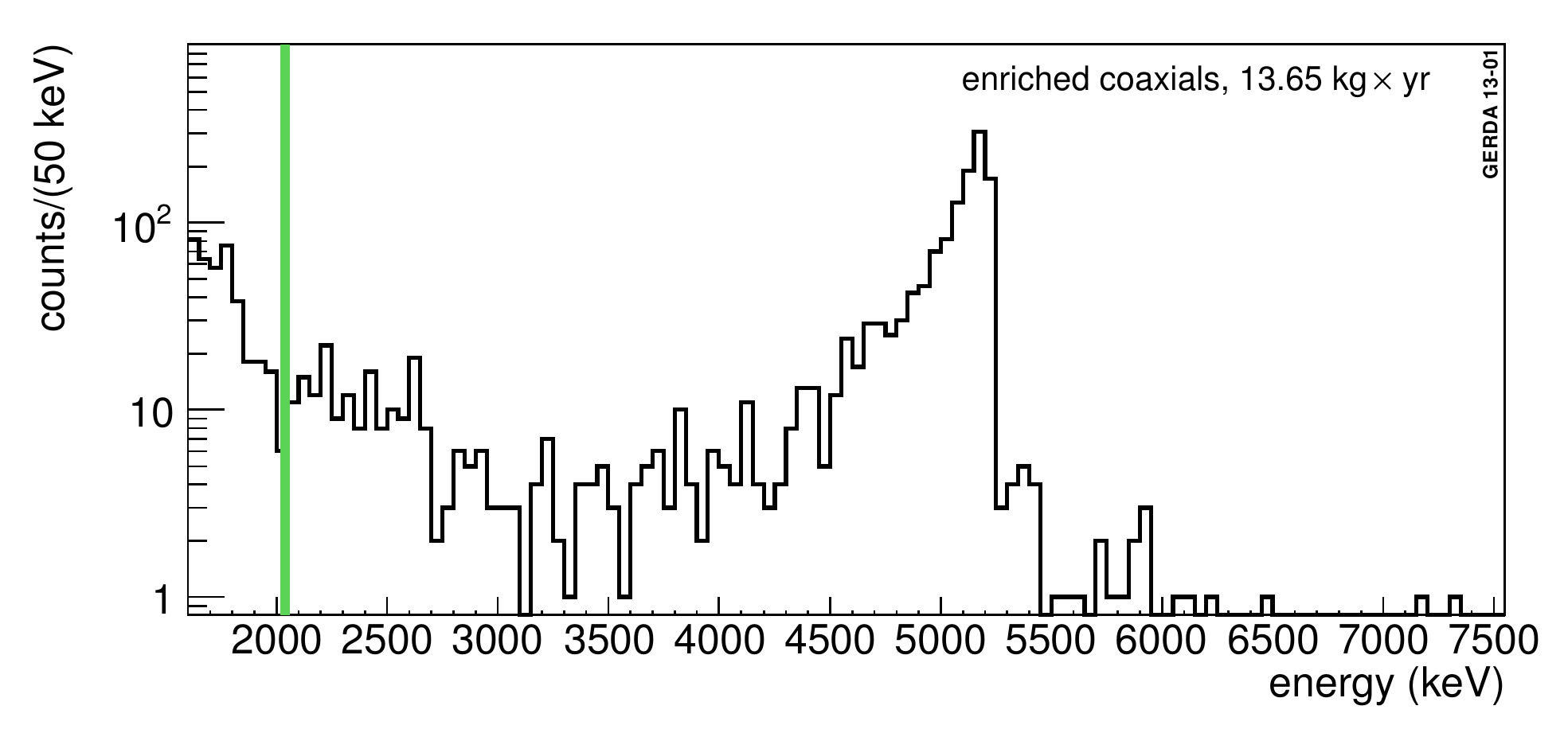}
\caption{ Energy spectrum of the \enrgecoax\ in high energy region measured in \gerda\ Phase I.
 }\label{fig:highenergy}
\end{figure}

A very prominent peak structure around 5.3~MeV with a tail towards lower energies 
has been observed in the energy spectrum of the \enrgecoax\ 
(depicted in Figure~\ref{fig:highenergy}) due to $^{210}$Po $\alpha$ decays.
Also a significant number of events observed above 5.3~MeV reveals that 
there are other sources than $^{210}$Po contributing to the spectrum. 
Other peak structures observed with lower intensities, i.e., around 4.7 MeV, 5.4 MeV and 5.9 MeV, 
indicate a contribution from the successive $\alpha$ decays in $^{226}$Ra decay chain. 

$\alpha$ particles with energies between 4\,MeV and 9\,MeV have several tens of $\mu m$ range in Ge and in LAr. Therefore, they can only deposit energy in the active volume 
if they decay on or close to the p$^{+}$ surface ($dl$ < 1\,$\mu m$) of the detectors and can induce events at \qbb\ 
after their energies degrade in LAr and $dl$.

In the following, analyses of the event rate distributions and of the energy spectrum 
of events above 3.5~MeV are described. While the source of the events which are 
dominant in different regions can be inferred from the time analysis, a model of the 
energy spectrum can allow to estimate their contributions around \qbb.     
\subsubsection{Bayesian inference} 
 The probability of the model and its
 parameters, the posterior probability is given from Bayes theorem as
\begin{equation}
\label{eq:posterior} 
P(\vec{\lambda}|\vec{n}) = \frac{P(\vec{n}|\vec{\lambda}) P_{0}(\vec{\lambda})}{\int P(\vec{n}|\vec{\lambda}) P_{0}(\vec{\lambda}) d\vec{\lambda}} 
\end{equation}
 where P($\vec{n}|\vec{\lambda}$) denotes the likelihood and
 P$_{0}(\vec{\lambda})$ the prior probability of the parameters. 
 When analyzing the binned distributions of data arising from a Poisson process, the
 likelihood is written as the product of the probability of data given the
 model and parameters in each bin
\begin{equation} 
P(\vec{n}|\vec{\lambda}) = \prod_{i} P(n_{i}|\lambda_{i})
= \prod_{i} \frac{e^{-\lambda_{i}} \lambda_{i}^{n_{i}} }{ n_{i}! }
\end{equation} 
 where $n_{i}$ is the observed number of events and $\lambda_{i}$ is the
 expected number of events in the $i$-bin bin.

 The analysis of both event rate and energy distributions is carried out by
 fitting the binned distributions due to the method described. One of the merits of 
 Bayesian analysis is the possibility to add initial 
 knowledge to the analysis in order to get a better answer. This is done by giving prior probabilities 
 on the parameters whenever available.
 The computation is done using the Bayesian Analysis Toolkit BAT~\cite{bat}.

\subsubsection{Event rate analysis}
The event rate distributions are obtained for two different energy regions;
3.5 -- 5.3~MeV where $^{210}$Po is dominant and 5.3--7.5~MeV where events only due to 
$^{226}$Ra decay chain are expected.
The distribution for the first region follows an exponential decrease as expected from 
an initial $^{210}$Po ($T_{1/2}=138.4$\,d) contamination. Two models are used to fit the distribution; 
an exponential rate (assuming only $^{210}$Po) 
and exponential plus a constant rate (allowing other contributions) by giving a Gaussian prior on the 
half life parameter with a mean value of 138.4 days and a standard deviation of 0.2 days. 
Note that a constant component can be due to $^{210}$Pb ($T_{1/2}=22.3$\,yr) and/or $^{226}$Ra ($T_{1/2}=1600$\,yr). 
The distribution for the second region looks flat as expected from $^{226}$Ra and fitted only with a constant. 

The expected number of events, $\lambda_{i}$, is corrected for the live time fraction. 
 While performing a fit with an exponential function it is written as
\begin{equation} \lambda_{i}
 = \epsilon_{i} \int_{(i-1)\Delta{t}}^{i\Delta{t}} \!
 { N_{0} \cdot e^{-\ln2 \, {t}/T_{1/2}}\,dt } \end{equation}  
 where $\epsilon_{i}$ is the value in the $i$-bin bin of the live time fraction
 distribution, $\Delta{t}$ is the bin width, $N_{0}$, the initial event rate
 and $T_{1/2}$, the half life are the parameters of the model.
\begin{figure}[htpb]
\includegraphics[width=0.40\columnwidth]{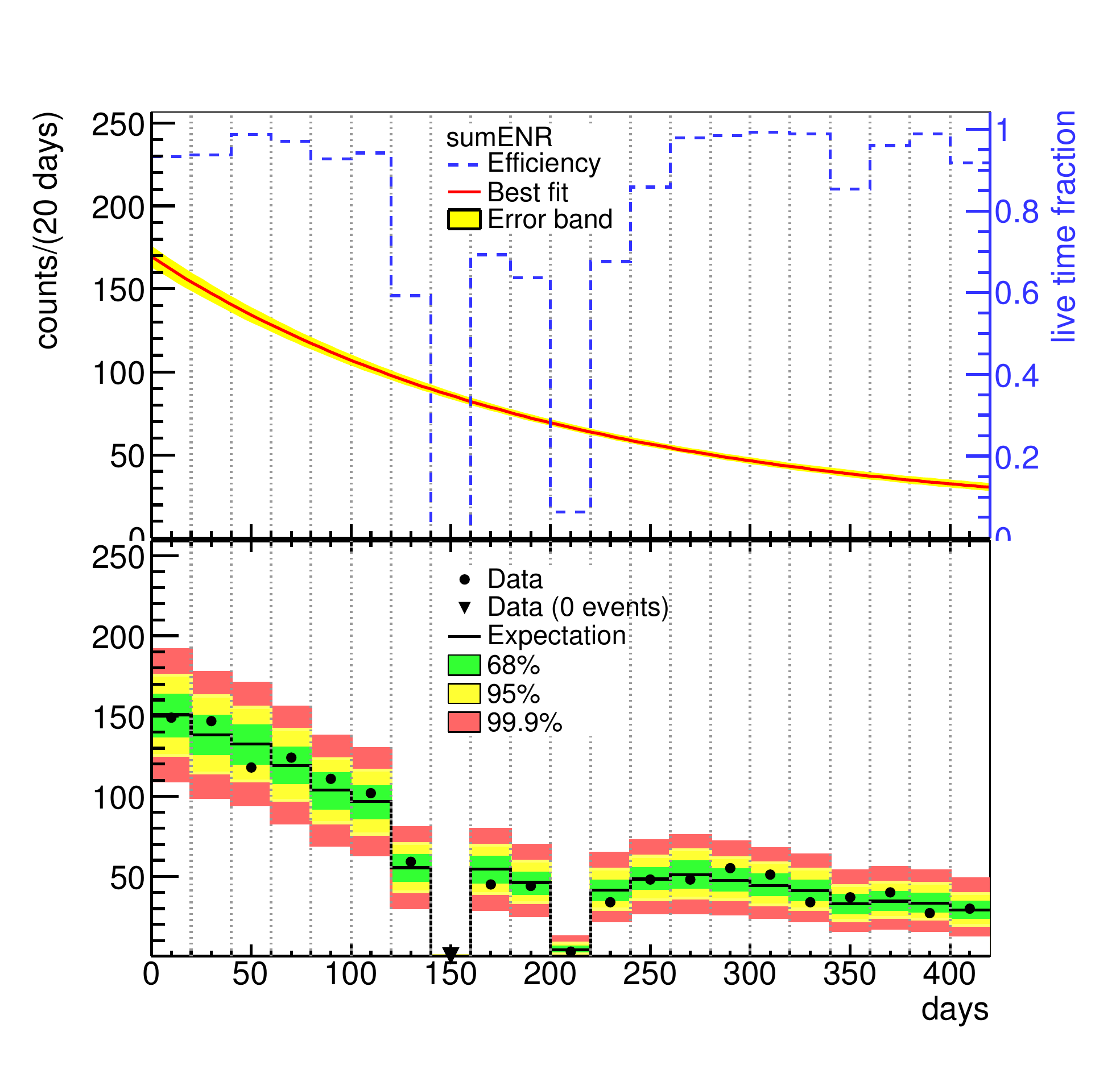}
\includegraphics[width=0.40\columnwidth]{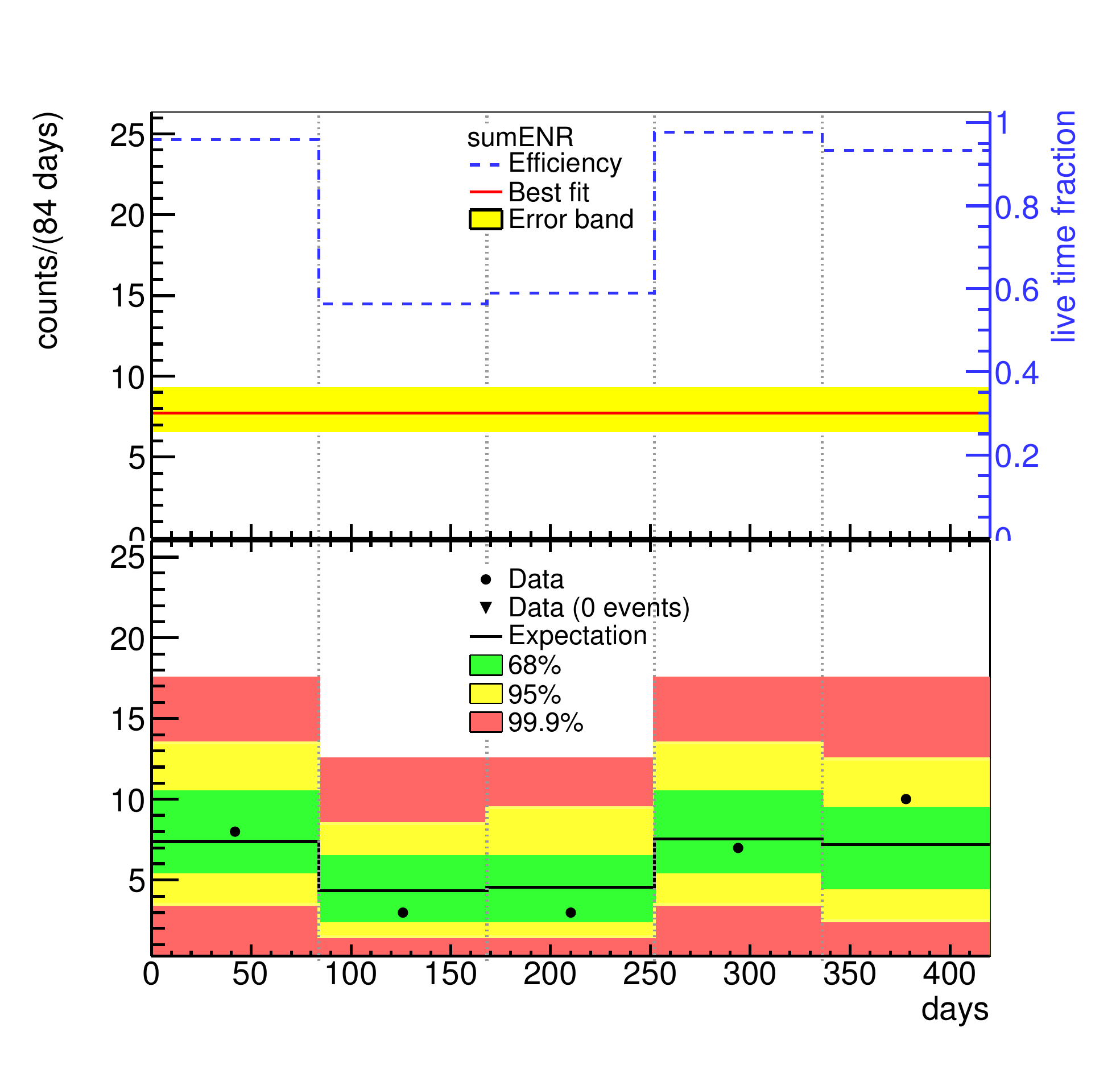}
\caption{ Results of fitting the event rate distribution for the 3.5--5.3\,MeV region with an exponential plus constant
(left) and for the 5.3--7.5\,MeV region with a constant. The best fit model with 68\% uncertainty band and the live time fraction distribution are shown in the upper panels. The observed and the expected number of events in the lower ones. Also shown are the smallest intervals containing the 68\%, 95\% and 99.9\% probability for the expectation in green, yellow and red regions, respectively~\cite{colorplots}. 
}\label{fig:eventrate}
\end{figure}
The distribution for the 3.5 -- 5.3~MeV region was described better with the second model with
a small constant term of (0.57$\pm$0.16) cts/day to an initial rate of (7.9$\pm$0.4) cts/day decreasing
exponentially with a half life of (138.4$\pm$0.2) days. A fit with a non informative prior on the half 
life parameter was also performed, resulting in a half life of (130.4$\pm$22.4) days, which is in very 
good agreement with the half life of $^{210}$Po. 
The distribution for the 5.3 -- 7.5~MeV region is described very well with a constant rate of 
(0.09$\pm$0.02) cts/day, supporting the assumption of an initial $^{226}$Ra source.

Figure~\ref{fig:eventrate} shows the results of the performed fits, i.e. data (what has been measured)
together with the expectation (what is expected to be measured given the live time fraction) due to   
the best fit model (what was supposed to be measured for a 100\% live time fraction). 
Comparison of data and expectation due to model is done by giving 
68\%, 95\% and 99.9\% probability intervals for the expectation.
\subsubsection{Spectral analysis}
Analysis of the energy spectrum is done under the assumption that events above 3.5\,MeV 
come from $^{210}$Po $\alpha$ decays and from the successive $\alpha$ decays in the $^{226}$Ra decay chain. 
While former is only assumed on the p$^{+}$ surface, the latter starting from $^{222}$Rn assumed also in LAr. 
This is expected since $^{222}$Rn emanates into LAr from materials with $^{226}$Ra contamination in 
the close vicinity of the detectors. 
The expected energy spectrum of all the components are obtained through MC 
simulations in \mage~\cite{mage} by using a detailed description of the \gerda\ Phase I setup. Spectra for different
$dl$ thicknesses (100\,nm -- 1\,$\mu m$) are simulated to derive the effective $dl$ thickness.   

\begin{figure}[htpb]
\includegraphics[width=0.53\columnwidth]{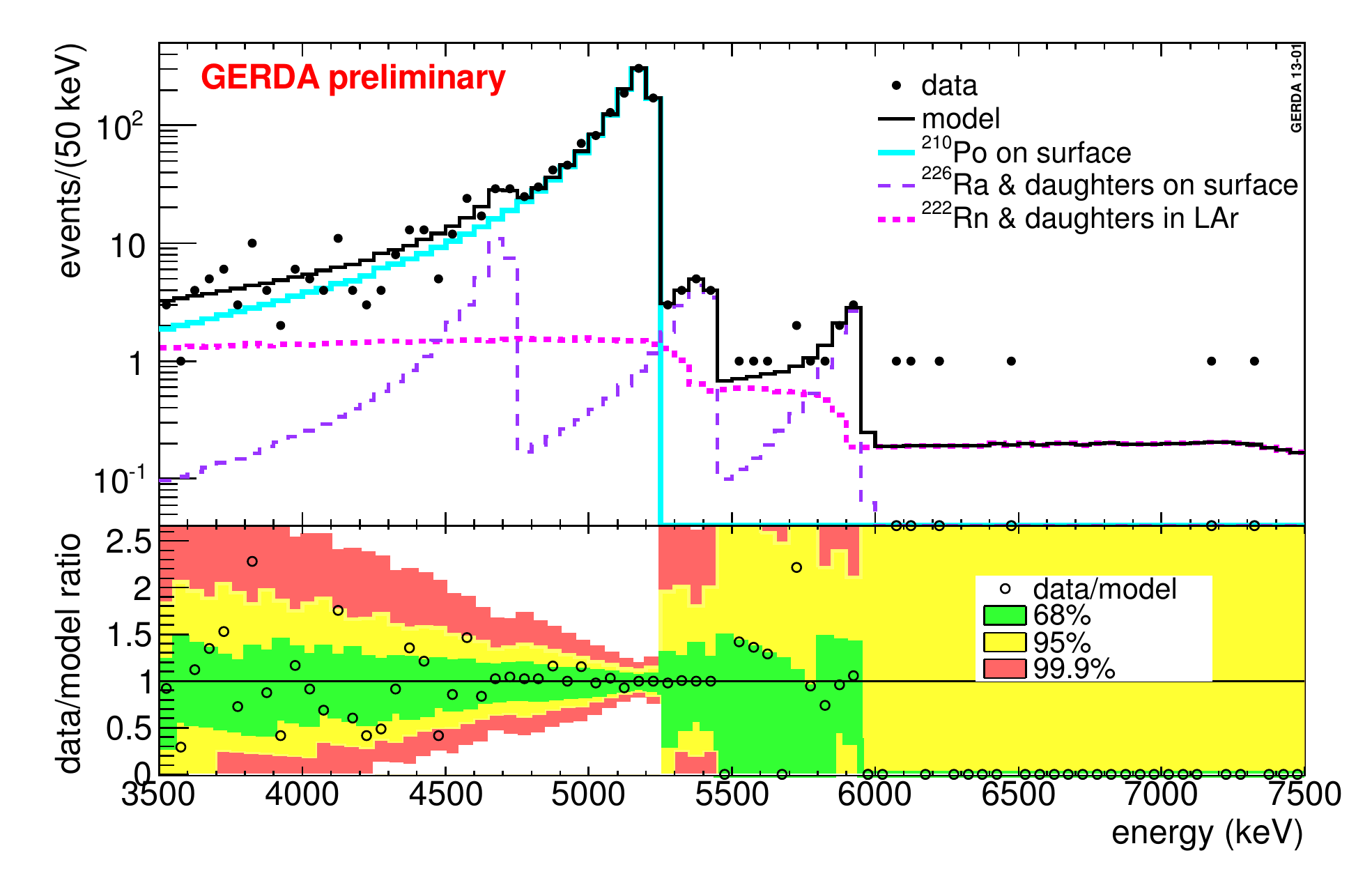}
\includegraphics[width=0.53\columnwidth]{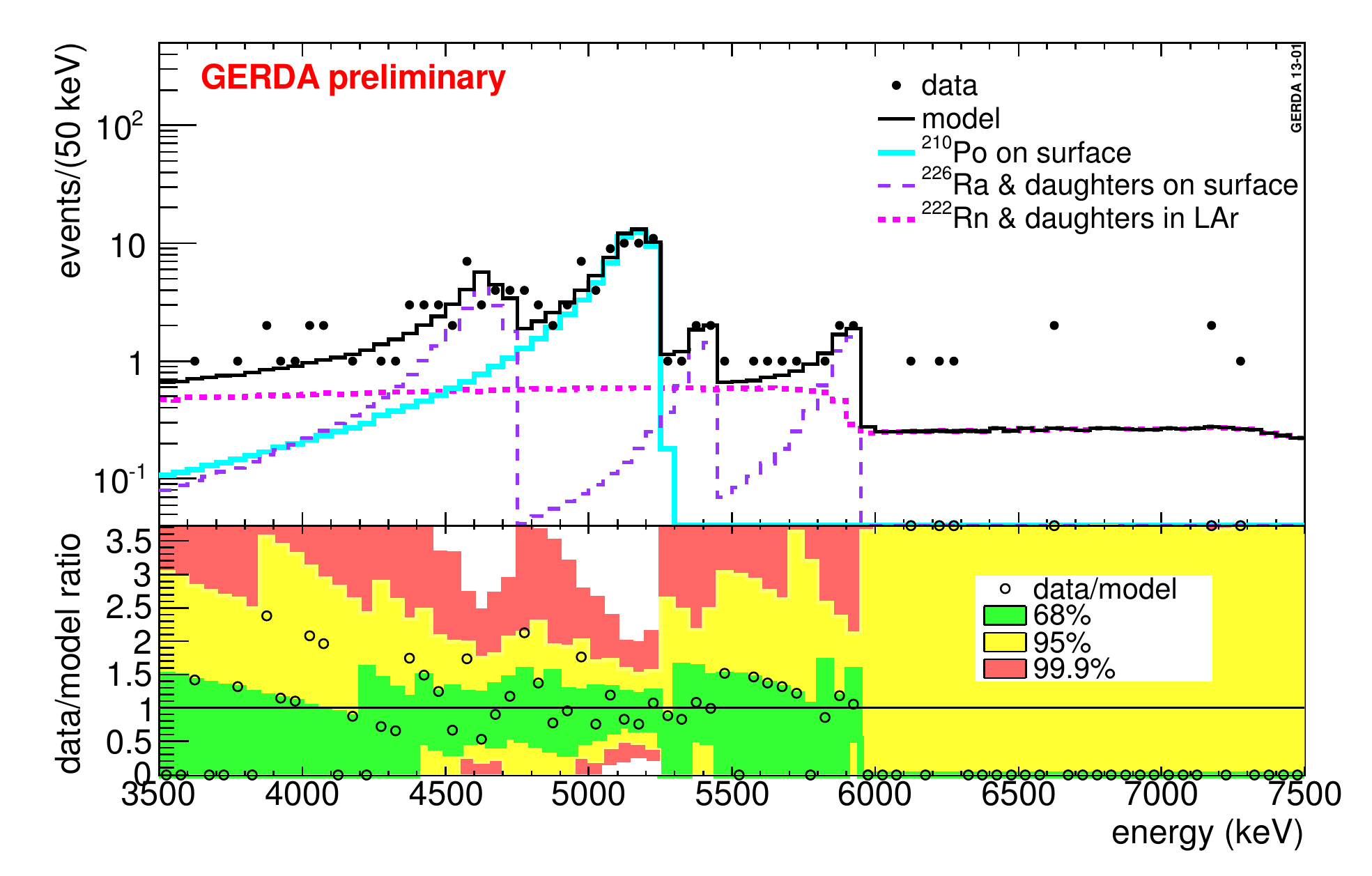}
\caption{ The upper panels show the best fit model (black histogram) and observed spectrum
(black markers) for \enrgecoax\ (left) and \natgecoax (right).
Individual components of the model are shown as well. The lower panel shows the ratio of data
and model and the smallest intervals containing 68\%, 95\% and 99.9\% probability for the model expectation.
  }\label{fig:alphafits}
\end{figure}

The simulated spectra are fitted to the observed spectrum with
a 50\,keV binning in 3.5--7.5\,MeV region by giving flat priors on the parameters. 
The analysis is also done for the \natgecoax\, which shows a similar spectral features but a lower $^{210}$Po rate 
and enhanced structures from $^{226}$Ra decay chain. The model describes both spectra very well 
(see Figure~\ref{fig:alphafits}). The results
are stable wrt. the choice of bin width. According to the model the expected background contribution in \qbb$\pm$5\,keV is
(2.1$\pm$0.5)$\cdot$10$^{-3}$\,\ctsper\ ($\sim$10\%)  for the \enrgecoax\, mostly (7\%) coming from LAr decays resulting 
in a linear spectrum with a small slope unlike surface decays. The contribution of $\alpha$ induced events depends on the
analyzed data set, since the surface contaminations are detector dependent and initial $^{210}$Po rate is decreasing in time.

\subsection{A global model for the background spectrum}
\begin{figure}[htpb]
\includegraphics[width=0.55\columnwidth]{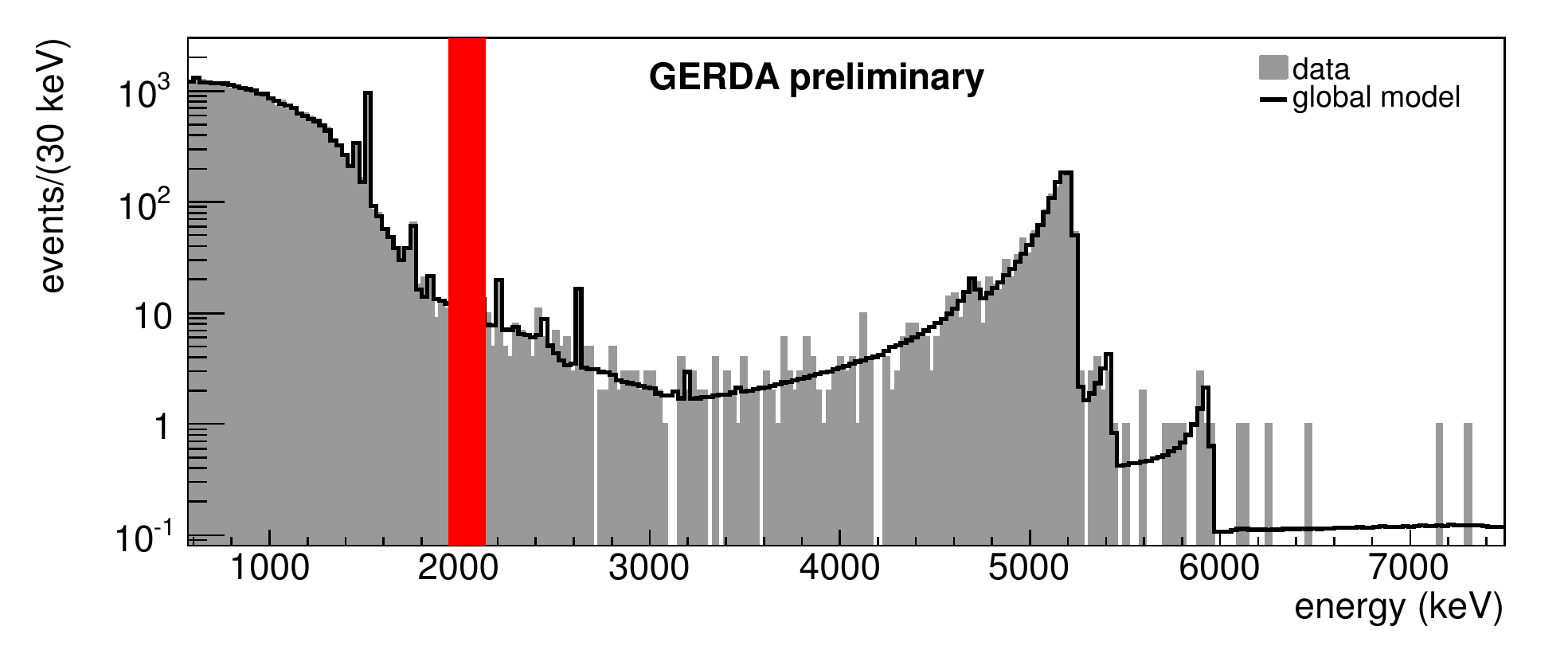}
\caption{Data (filled histogram) from the enriched coaxial detectors and the best fit model (black histogram). The red band
masks the region of interest. }\label{fig:fullspectrum}
\end{figure}

The $\alpha$ induced event model alone successfully describes the observed energy spectrum 
down to 3.5~MeV. Below 3.5~MeV many other background components contribute 
to the spectrum, some of them also relevant around \qbb. 
The analysis window is therefore expanded down to 570~keV to obtain a full background decomposition at \qbb. 
The part where the beta spectrum of $^{39}$Ar (Q$_\beta$=565\,keV) is the dominating component without any 
relevance at \qbb\ is not included in the analysis. 
All the components -- namely, 2$\nu\beta\beta$ decay of $^{76}$Ge, 
$^{42}$K, $^{40}$K, $^{214}$Bi, $^{228}$Th, $^{60}$Co and the $\alpha$ model --  
that are expected to contribute in this energy window are considered in a global fit.
Some parameters are given an informative prior probability, 
e.g. a Gaussian prior probability distribution for the expected $^{214}$Bi decays on the p$^{+}$ surface 
is given according to the $\alpha$ model.
The best fit model together with the observed spectrum is shown in
Figure~\ref{fig:fullspectrum}, which is rather a qualitative demonstration of the 
success of the model. Many cross-checks and systematic uncertainties are still under study. 
Therefore, the details of the analysis and its results are intentionally neither shown nor discussed. 
Nevertheless, one important conclusion can be made: The main contributions around \qbb\ come 
from $^{214}$Bi, $^{228}$Th, $^{42}$K, $^{60}$Co and $\alpha$ emitting isotopes in the $^{226}$Ra decay chain, with a fraction depending on the assumed source position and distribution. 

\section{Conclusions}

A model for the observed background energy spectrum in \gerda\ Phase I is obtained. 
The model allows to have a decomposition of the background at \qbb.
Other important informations that can be deduced from the background model are, the expected number 
of background events and the spectral shape of background around \qbb. These are essential inputs for a reliable
result in the upcoming \nonu\ analysis. After all the necessary cross checks are performed and 
systematic uncertainties are evaluated, the results will be presented in a paper from the \gerda\ collaboration.

\bibliographystyle{aipproc}   

\bibliography{sample}

\IfFileExists{\jobname.bbl}{}
 {\typeout{}
  \typeout{******************************************}
  \typeout{** Please run "bibtex \jobname" to optain}
  \typeout{** the bibliography and then re-run LaTeX}
  \typeout{** twice to fix the references!}
  \typeout{******************************************}
  \typeout{}
 }

\end{document}